\documentclass{article}
\usepackage{dcase2020_techrep,amsmath,graphicx,url,times,booktabs, tabularx}

\usepackage{hyperref}
\usepackage{xspace} 

\title{CRNNs for urban sound tagging with spatiotemporal context}

\twoauthors
  {Augustin Arnault\sthanks{Work done as part of Multitel internship.}}
    {IMT Lille Douai \\
    Lille, France \\
     augustin.a@free.fr}
  {Nicolas Riche}
    {Multitel \\
     Mons, Belgium \\
     riche@multitel.be}

\begin{document}

\ninept
\maketitle

\begin{sloppy}

\begin{abstract}
This paper describes CRNNs we used to participate in Task 5 of the DCASE 2020 challenge. This task focuses on hierarchical multilabel urban sound tagging with spatiotemporal context. The code is available on our GitHub repository at \href{https://github.com/multitel-ai/urban-sound-tagging}{https://github.com/multitel-ai/urban-sound-tagging}.
\end{abstract}


\section{Introduction}
\label{sec:intro}

This paper describes our submission to the urban sound tagging challenge, which is carried out as Task 5 (see Fig. \ref{task}) of the DCASE 2020 challenge. We built an urban sound tagging neural network which takes as input (log-mel) spectrograms and metadata (week, day, hour and location) and outputs mulitlabel prediction vector. There are two different levels of granularity as shown in Fig. \ref{taxonomy}. The first one returns whether each of 23 sources of noise (fine-grained tags) is audible in the recording or not. The second one predicts coarse-grained tags among a list of eight. The relationship between coarse-grained and fine-grained tags is hierarchical. It is, therefore, possible to derive coarse-grained labeling from fine-grained labeling, but not the other way around. \\

\begin{figure}[t]
  \centering
  \centerline{\includegraphics[width=\columnwidth]{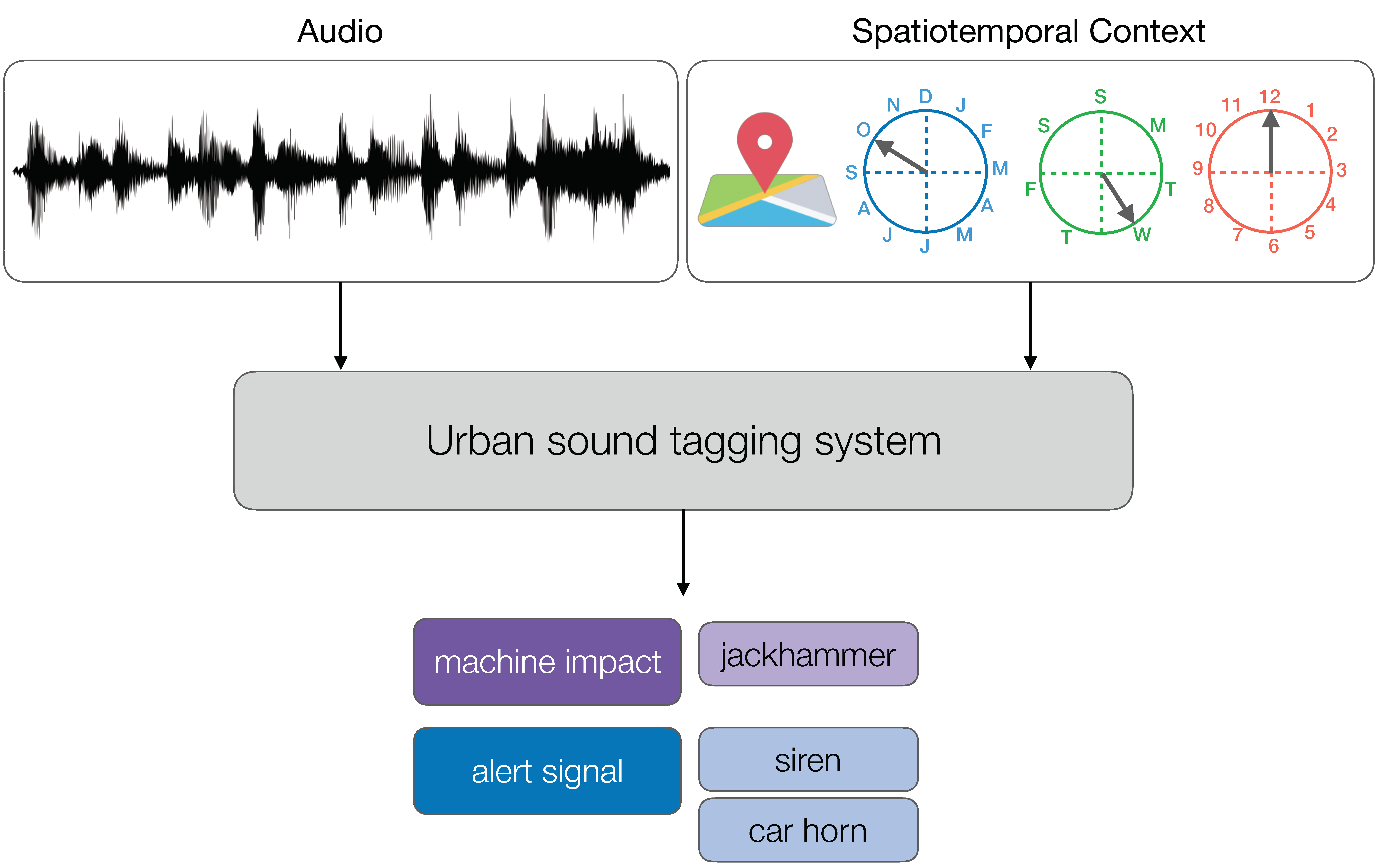}}
  \caption{Task 5 of DCASE 2020 \cite{dcase}.}
  \label{task}
\end{figure}

\begin{figure}[t]
  \centering
  \centerline{\includegraphics[width=\columnwidth]{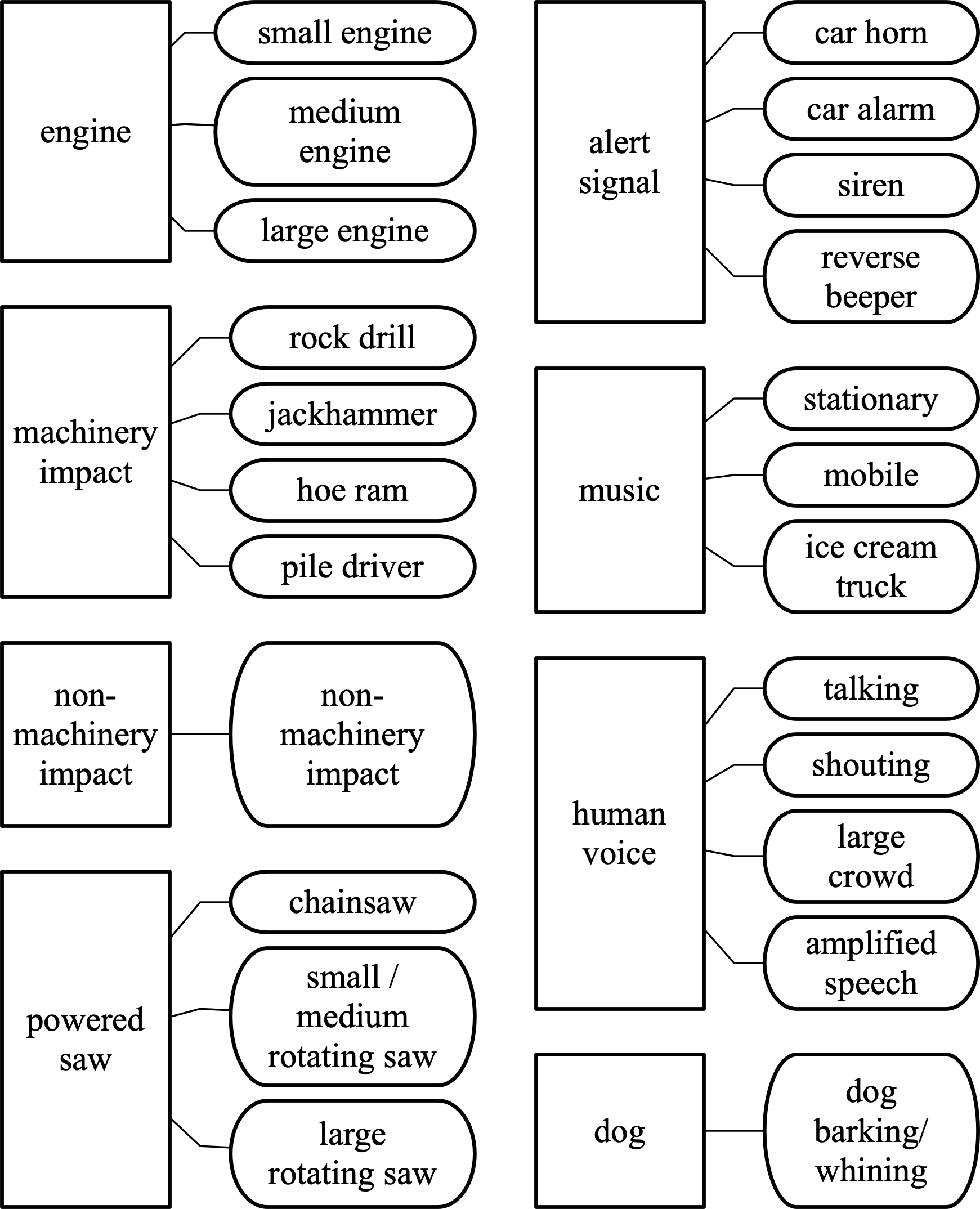}}
  \caption{Hierarchical taxonomy of urban sound tags in the Task 5 of DCASE 2020.}
  \label{taxonomy}
\end{figure}

Since 2013, the DCASE \footnote{Detection and Classification of Acoustic Scenes and Events} challenges have provided numerous publicly available datasets and have gained an increasing research interest in audio pattern recognition. Though, there is still a need for a large scale dataset of generic real-world sound like ImageNet in image classification or Wikipedia data in natural language processing. To address this issue, in 2017, Google released AudioSet \cite{audioset}. This dataset contains 2.1 millions of 10s audio sound grabbed from YouTube videos and annotated with presence / absence labeling of 527 types of sound events. \\

The DCASE task 5 uses the SONYC-UST \footnote{SONYC Urban Sound Tagging} \cite{dcase} as main dataset. Similar to Audioset, it provides about 17 thousand 10s samples with coarse-grained and fine-grained tags (see Fig. \ref{taxonomy}) alongside a diversity of metadata such as spatio-temporal context. This dataset has been recorded by the SONYC acoustic sensor network and tagged by volunteers and SONYC team members.

%%%%%%%%%%%%%%%%%%%%%%%%%%%%%%%%%%%%%%%%%%%%%%%%%%%%%%%%%%%%%%%%%%%%%%%%%%%%%%%%

\section{RELATED WORK}
Deep neural networks (DNN) have been extensively studied and applied over the last decade. Several deep learning neural networks have been proposed in machine listening research. First, as mentioned and summarized in \cite{panns}, several CNN-based architectures have been applied to (log-mel) spectrogram of audio recordings followed by an activation function to predict the presence or absence of sounds. \\

Although CNNs act as a robust feature extractor, their receptive field have limited size. Therefore, they cannot capture long time dependency. To solve this problem, CRNNs were proposed to consider the long time information. The idea is to use a CNN combined with RNN or RNN-like layers such as GRU or LSTM. For example, in \cite{talnet} a CNN followed by GRU layer is used. Moreover, this model includes a final pooling layer as it was designed for MIL \footnote{Multiple Instance Learning} problems. Extensive comparison of pooling strategies has been made in \cite{talnet-pooling}.\\

Last but not least, as explained in \cite{cnn-transfo}, transformer has been proposed to take into account the long time dependency of time series. This approach is inspired of the "Attention Is All You Need" work and the success it acquired in the NLP field \cite{attention}. Replacing RNN layers by self attention layers has been proven to increase performances in general.

%%%%%%%%%%%%%%%%%%%%%%%%%%%%%%%%%%%%%%%%%%%%%%%%%%%%%%%%%%%%%%%%%%%%%%%%%%%%%%%%

\section{PREPROCESSING AND DATA AUGMENTATION}
\subsection{Spectrogram generation}
Recordings are resampled to 44100 Hz and to generate (log-mel) spectrogram as the representation for the audio input data. Librosa \cite{librosa} was used to compute these (log-mel) spectrograms. To compute and transforme the STFT to (log-mel) spectrogram, a Hanning window size of 2822 and hop length of 1103 samples were used with the number of bands being 64. The frequencies between 0Hz and 8000Hz are kept. Those numbers has been chosen to match those of TALNet original implementation \cite{talnet}.

\subsection{Data augmentation}
We used SpecAugment \cite{spectaugment} which  consists  of  warping  the features, masking blocks of frequency channels, and masking blocks of time steps to supplement the training data. Moreover, several image data augmentation techniques \cite{da} were employed such as ShiftScaleRotate, Grid distortion and Cutout. Mixup \cite{mixup}, a method that linearly mixes two random training examples with a scalar lambda sampled from a beta distribution, was used as well. We found out it was helping the model to obtain better scores.

\subsection{Relabeling}
A relabelling strategy has been used to relabel all the training set and a part of the validation set (the ~4000 samples not labelled by the SONYCUST team). The ~500 samples annotated by SONYCUST team remain untouched. The model used to relabel is System 2 (see below in Section \ref{archi}) saved after reaching maximum macro AUPRC on coarse. Then, the 3 systems learn on the relabeled dataset.
%%%%%%%%%%%%%%%%%%%%%%%%%%%%%%%%%%%%%%%%%%%%%%%%%%%%%%%%%%%%%%%%%%%%%%%%%%%%%%%%
\section{FEATURE REPRESENTATION}

\subsection{Generic audio embeddings}
One solution to generate generic audioset embeddings is to use released embeddings of audio clips extracted from a frozen neural network as feature extractor like OpenL3 \cite{openl3} or VGGish \cite{vggish}.
However,  as explained in \cite{panns}, those methods did not work on improving systems obtaining better embedding features. Therefore, instead of using them, we decided to work on transfer learning. \\

One architecture have been extensively tested for this DCASE challenge: TALNet \cite{talnet}. All parameters to calculate the embedding features are initialized from the pretrained audioset weights instead of being randomly initialized.

\subsection{Specific audio embeddings}
\label{transfo_ex}
We trained a neural network from scratch to generate specific DCASE embeddings. A TALNet-like architecture has been improved. All parameters are randomly initialized. \\

Three improvements have been incorporated into the neural network.  First, Group Normalization (GN) \cite{gn} is used instead of Batch Normalization (BN) to be independent of the batch dimension.  Then, a second normalization technique called Weight Standardization \cite{ws} is used to accelerate training and smooth the loss and the gradients. Finally, the bi-GRU layer has been replaced by an encoder layer of a transformer \footnote{The decoder part which transforms an embedding back to output is not used.} to decrease the number of parameters and increase performances. Each encoder consists of several encoder layers. For each layer, query, key and value transform matrices (see in \cite{attention}) were used on the outputs of the last convolution. After the computation of the feature correlation of different time steps, a softmax operation converts the correlation value to probability along the time steps. 

\begin{figure}[h]
  \includegraphics[width=\columnwidth,height=4cm]{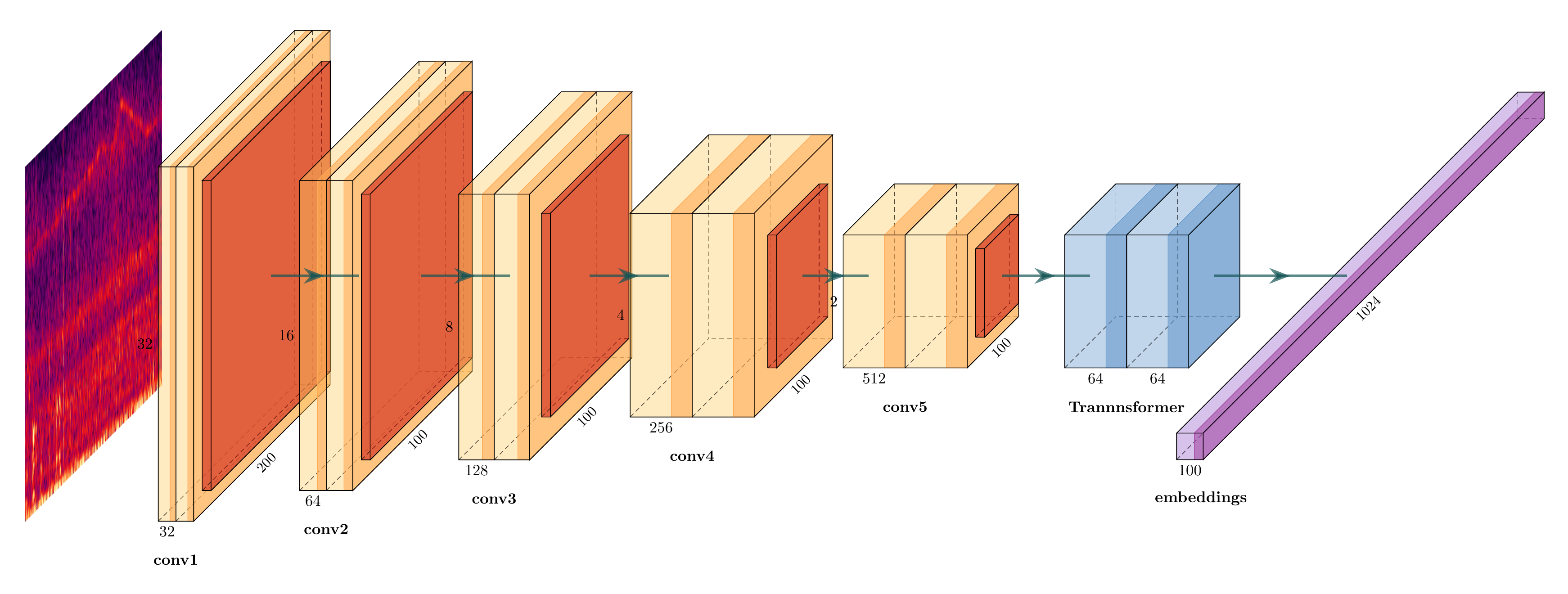}
  \caption{Overview of our TALNet-like architecture to compute specific audio embeddings.}
  \label{UrbanNet}
\end{figure}

\subsection{Metadata embeddings}
In \cite{t2v}, the author provides a model-agnostic vector representation for time, called Time2Vec, that can be easily imported into many existing and future architectures and improve their performances. The T2V representation was applied to all cyclic metadata because the representation is simple, invariant to re-scaling and captures both periodic and non-periodic patterns. \\

\begin{figure}[h]
  \centering
  \centerline{\includegraphics[width=\columnwidth]{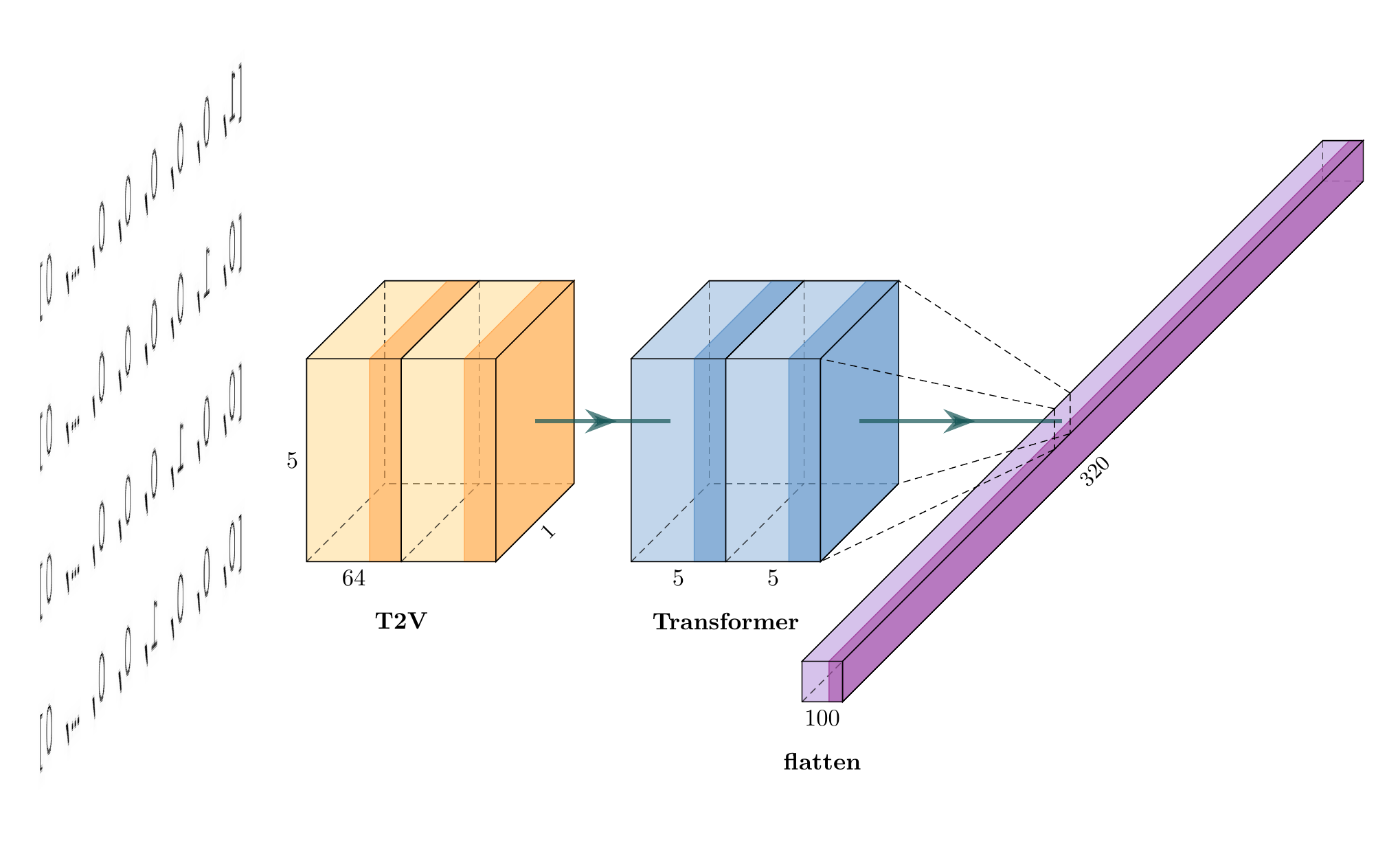}}
  \caption{The architecture of metadata embeddings.}
  \label{meta_emb}
\end{figure}

The encoder part of a transformer (see Section \ref{transfo_ex}) is then applied to transform the T2V representation to a high level embedding.

%%%%%%%%%%%%%%%%%%%%%%%%%%%%%%%%%%%%%%%%%%%%%%%%%%%%%%%%%%%%%%%%%%%%%%%%%%%%%%%%
\section{MODELS}
\label{archi}

\subsection{System1: Specific + Metadata}
The first architecture takes as input the (log-mel) spectrogram and the metadata to generate two embeddings. First, a specific audio embedding is generated using the improved version of TALNet network. Then, the metadata embeddings are created by a T2V-Transformer network. Finally, the last layer is a fully-connected layer which converts these two embeddings (concatenated) into a multi-label classification problem. The neural network has been trained to label both coarse and fine tags jointly. System1 outputs a prediction vector of 31 sources of noise (8 coarse-grained tags + 23 fine-grained tags).

\subsection{Systems 2 and 3: Generic + Specific + Metadata}
As you can see in Fig. \ref{archi_fig}, the architecture of Systems 2 and 3 take as input the (log-mel) spectrogram and the metadata to generate three embeddings.  First, a specific embedding is created by a TALNet-inspired CNN-Transformer. Then, a generic embedding is generated by a pre-trained TalNet network. Finally, the metadata embeddings are created by a T2V-Transformer network. The last layer is a fully-connected layer which converts these three embeddings (concatenated) into a multi-label classification problem. \\

\begin{figure}[h]
  \centering
  \centerline{\includegraphics[width=\columnwidth]{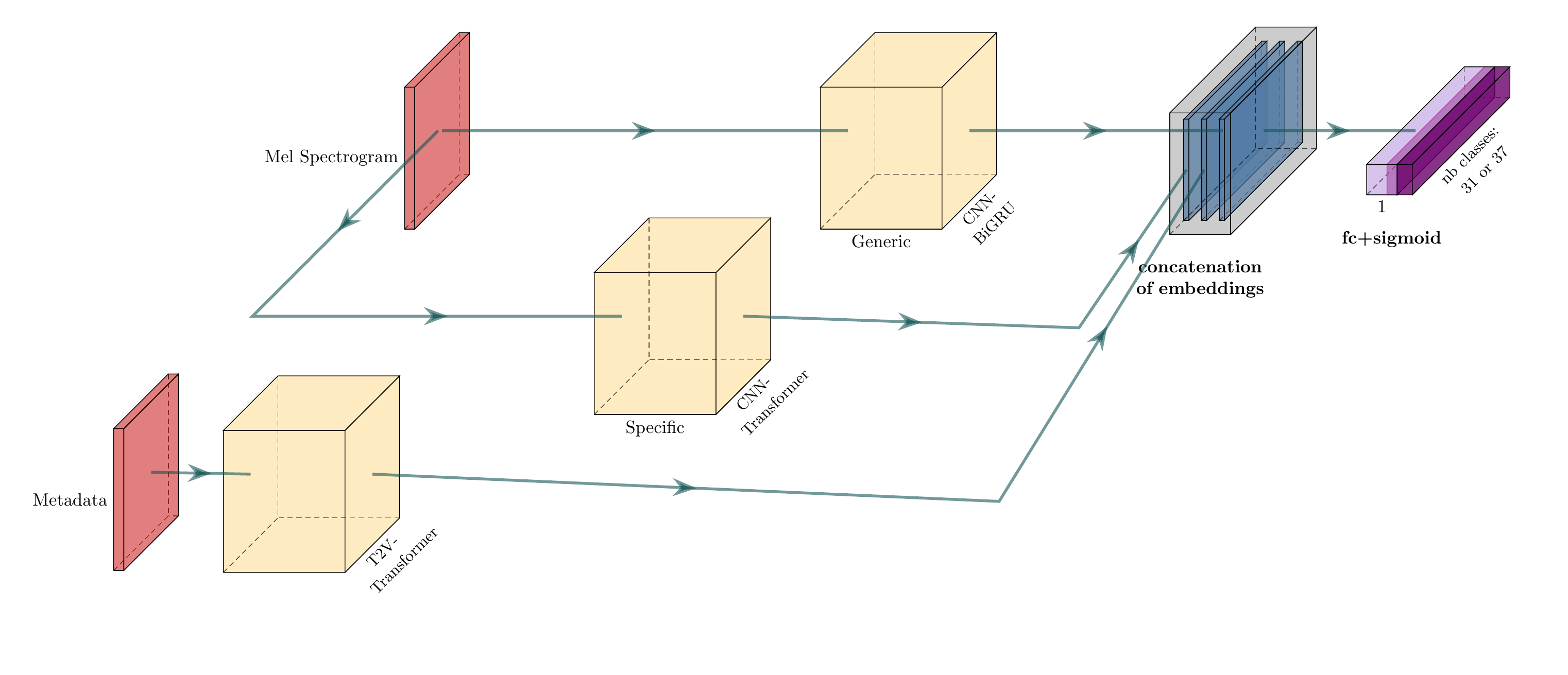}}
  \caption{The structure of Systems 2 and 3 for urban sound tagging with spatiotemporal context.}
  \label{archi_fig}
\end{figure}

The neural network has been trained to label both coarse and fine tags jointly. Two Systems have been submitted: System2 outputs a prediction vector of 31 sources of noise (8 coarse-grained tags + 23 fine-grained tags). System3 outputs 37 sources of noise. The fine other/unknown classes have been included during the training.

%%%%%%%%%%%%%%%%%%%%%%%%%%%%%%%%%%%%%%%%%%%%%%%%%%%%%%%%%%%%%%%%%%%%%%%%%%%%%%%%
\section{TRAINING TECHNIQUES}
Training was done on PyTorch Lightning \cite{pl}, Ralamb (Radam \cite{radam} + LARS  \cite{lars}) variant of the Adam algorithm with a learning rate of $10^{-3}$ was used as standard optimizer. On top of Ralamb, we use an algorithm called Lookahead which chooses a search direction, computes weight updates by looking ahead at the sequence of “fast weights" generated by the Ralamb optimizer. In \cite{lookahead}, the author show that Lookahead improves the learning stability and lowers the variance of its inner optimizer with negligible computation and memory cost. \\

The Systems 1 and 2 presented in Section \ref{archi} were trained with a jointly loss: Binary cross entropy (BCE) was used for the coarse-level and a Masked BCE loss (see in \cite{dcase}) was used for the fine-level. System 3 outputs 37 predictions and a unique BCE loss was used for training.

%%%%%%%%%%%%%%%%%%%%%%%%%%%%%%%%%%%%%%%%%%%%%%%%%%%%%%%%%%%%%%%%%%%%%%%%%%%%%%%%
\section{RESULTS} 
We evaluate our models on the provided validation dataset. As the primary classification metric, the challenge uses the macro-averaged Area Under the Precision-Recall Curve (macro-averaged AUPRC) for ranking. \\

Our results can be found and compared to the baseline in Table \ref{results_c} for coarse level and in Table \ref{results_f} for fine level. \\

\begin{table}[h]
\begin{center}
\caption{Results for coarse level}
\label{results_c}
\begin{tabular}{llll}\toprule
& \multicolumn{3}{c}{$\textbf{COARSE-GRAINED}$}
\\\cmidrule(lr){2-4}
           &Micro  & Micro & Macro\\
           & AUPRC  & F1    & AUPRC\\\midrule
Baseline& 0.8352 & 0.7389 & 0.6323 \\
System1 & 0.8622  & 0.7730 & 0.7601 \\
System2 & 0.8906  & 0.7953 & 0.8011 \\
System3 & \textbf{0.8956}  & \textbf{0.8039} & \textbf{0.8107}  \\\bottomrule
\end{tabular}
\end{center}
\end{table}

\begin{table}[h]
\begin{center}
\caption{Results for fine level}
\label{results_f}
\begin{tabular}{llll}\toprule
& \multicolumn{3}{c}{$\textbf{FINE-GRAINED}$}
\\\cmidrule(lr){2-4}
           &Micro  & Micro & Macro\\
           & AUPRC  & F1    & AUPRC\\\midrule
Baseline  & 0.7329 &  0.6149 & 0.5278  \\
System1   & 0.7983 & 0.6788  & 0.6349  \\
System2   & 0.7932 & 0.6902  & 0.6817  \\
System3   & \textbf{0.8126} & \textbf{0.7116}  & \textbf{0.7040}  \\\bottomrule
\end{tabular}
\end{center}
\end{table}

Our methods were able to surpass the Micro and Macro AUPRC baseline scores in coarse-level and fine-level evaluation. 

\section{CONCLUSIONS}
This paper presents CRNNs for the Task 5 of the DCASE 2020. We investigated the performance of generic and/or specific audio embeddings with metadata embeddings. \\

In the future, we will continue to explore multilabel urban sound tagging, study the classwise performance and apply CRNNs in other tasks as sound event detection.

%%%%%%%%%%%%%%%%%%%%%%%%%%%%%%%%%%%%%%%%%%%%%%%%%%%%%%%%%%%%%%%%%%%%%%%%%%%%%%%%
\section{ACKNOWLEDGEMENT}
This research was carried out as part of the Wal-e-Cities project portfolio, co-financed the European Regional Development Fund (ERDF 2014-2020) and supported by the Walloon Government.

\addtolength{\textheight}{-12cm}   % This command serves to balance the column lengths

%%%%%%%%%%%%%%%%%%%%%%%%%%%%%%%%%%%%%%%%%%%%%%%%%%%%%%%%%%%%%%%%%%%%%%%%%%%%%%%%

% -------------------------------------------------------------------------
% Either list references using the bibliography style file IEEEtran.bst
%\bibliographystyle{IEEEtran}
%\bibliography{refs}
%
% or list them by yourself

\end{sloppy}
\end{document}